\documentclass[preprint,showpacs,showkeys,aps]{revtex4}
\usepackage{graphicx}

\begin{document}
\title{Smooth-Particle Phase Stability          \\
with density and density-gradient potentials     }
\author{Wm. G. Hoover and Carol G. Hoover       \\
Highway Contract 60, Box 565                    \\
Ruby Valley, Nevada 89833                       \\
}
\date{\today}

\pacs{62.10.+s, 62.20.-x, 62.20.Dc, 83.10.Mj, 83.10Ff}

\keywords{Smooth Particles, Stability, Density Gradients, Surface Tension}

\begin{abstract}
Stable fluid and solid particle phases are essential to the
simulation of continuum fluids and solids using Smooth
Particle Applied Mechanics.  We show that density-dependent
potentials, such as $\Phi _\rho = \frac{1}{2}\sum (\rho - \rho_0)^2$,
along with their corresponding constitutive
relations, provide a simple means for characterizing fluids
and that a special stabilization potential,
$\Phi _{\nabla \rho }= \frac{1}{2}\sum (\nabla \rho )^2$,
not only stabilizes crystalline solid phases (or meshes) but
also provides a surface tension which is missing in the usual
density-dependent-potential approach.  We illustrate these ideas
for two-dimensional square, triangular, and hexagonal lattices.
\end{abstract}

\maketitle

\section{Smooth Particle Applied Mechanics}
Smooth particle applied mechanics---SPAM---was discovered about
thirty years ago\cite{b1,b2}.  It has become a useful tool in
simulating gases, fluids, and solids, and holds particular
promise for problems involving large high-speed deformation and
failure.  The main advantage of the method is simplicity.  SPAM
closely resembles atomistic molecular dynamics and, in a variety
of cases\cite{b3}, the SPAM particle trajectories are isomorphic
to those of molecular dynamics.

A smooth-particle code is less-complicated than typical grid-based
continuum codes because the smooth-particle method evaluates
spatial gradients in a particularly simple way, explained in more
detail below.  The main disadvantages of the method are instability
in tension\cite{b4} and the lack of surface tension\cite{b5}.  The
present work introduces an idea---density-gradient
potentials---designed to address those problems.

The basic smooth-particle approach is to represent all continuum
properties (the density
$\rho $, the velocity $v$, the stress tensor $\sigma $, ...) as
interpolated sums of particle properties, where the particles
are described by ``weight functions'', expressing the range of
influence of the particles in space.  The {\em simplest} weight
function satisfying five desirable conditions---(i) normalization,
(ii) finite range $h$, (iii) a maximum at the origin, and
(iv and v) two continuous derivatives everywhere---is
Lucy's.  Normalized for applications in two-dimensional space Lucy's
weight function is as follows:
$$
w_{\rm Lucy}(r<h) = \frac{5}{\pi h^2}
\left[1 - 6\frac{r^2}{h^2} + 8\frac{r^3}{h^3} - 3\frac{r^4}{h^4}\right]
$$
$$
\longrightarrow \int _0^h 2\pi rw(r)dr \equiv 1 \ .
$$
The density at any point $r$ is defined as the sum of all the
particle contributions at that point:
$$
\rho(r) \equiv \sum _jm_jw(r-r_j) \ ,
$$
so that the density associated with Particle $i$ is
$$
\rho _i = \sum _jm_jw_{ij} \ ; \ w_{ij} \equiv w(|r_i-r_{j}|) =
w(r_{ij}) \ .
$$
Other continuum properties at location $r$ are likewise calculated
as sums over nearby particles:
$$
f(r)\rho (r) \equiv \sum _j m_jf_jw(|r-r_j|) \longrightarrow
$$
$$
f(r) = [f(r)\rho (r)]/\rho (r) =
$$
$$
\sum _j m_jf_jw(|r-r_j|)/\sum _jm_jw(|r-r_j|) \ .
$$
It is important to note that the interpolated function $f(r)$ at
the location of Particle $j$ is typically different to the particle
property $f_j$ at that point.

Beyond using the point properties $\{f_j \}$ to define the field properties
$f(r)$ this approach has the crucial advantage that {\em gradients}
of the field properties translate into simple sums of particle
quantities:
$$
\nabla (\rho f)_r = (f\nabla \rho )_r + (\rho \nabla f)_r \equiv 
$$
$$
\nabla _r\left[\sum _j m_jf_jw(|r-r_j|)\right] =
\sum _j m_jf_j\nabla _rw(|r-r_j|) \ .
$$
Expressions for the gradients of density, velocity, stress, and
energy make it possible to express the {\em partial} differential equations
of continuum mechanics as {\em ordinary} differential equations for the
evolution of the particle coordinates, velocities, stresses, and
energies\cite{b6}.  The
resulting ``equation of motion'' for the particles is
$$
\dot v_i = -\sum _j\left[(mP/\rho ^2)_i + (mP/\rho ^2)_j \right]\cdot
\nabla _iw_{ij} \ ,
$$
where $P_i$ is the pressure tensor associated with Particle $i$.  Where
the pressure is hydrostatic, and slowly varying in space, it is
noteworthy that the smooth-particle equations of motion are exactly
the same as the equations of molecular dynamics, with the weight
function $w(r)$ playing the r\^ole of a pair potential.  In
the simple case that the internal energy depends only on volume (and not on
temperature) the pressure is simply related to the internal energy per
unit mass $e$:
$$
P = \rho^2de/d\rho \ .
$$
Specifying the density dependence of either the pressure or the
internal energy, in such a case, corresponds to giving a full description
of the equilibrium equation of state.

\section{Convergence of Smooth-Particle Averages}
The evolving time-and-space dependent smooth-particle sums converge
to continuum mechanics as a many-particle limit, just as do the more
usual grid-based approximations.  The range of the smooth-particle
weight function, $h$, corresponds to a few grid spacings.  In
practice, for an error level of order one percent, the
weight-function sums must include a
few dozen particles.  Fig. 1 shows the dependence of density
sums, $\rho _i = \sum _jm_jw_{ij}$ on the range of the weight function for
three regular two-dimensional lattices.  In typical applications, with
$h \simeq 3\sqrt{V/N}$ the density errors are of order one percent.

\begin{figure}
\includegraphics{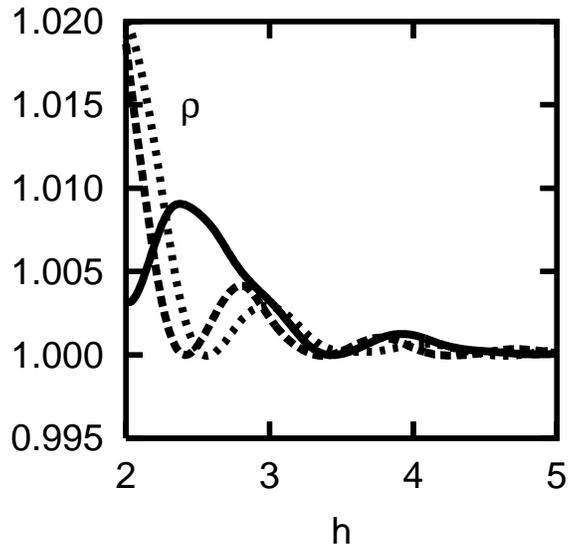}
\caption{
Summed-up densities evaluated {\em at} regular-lattice particle
sites for (from top to bottom at the left side of the plot)
the square, triangular, and hexagonal lattices.  The
range of Lucy's weight function varies from 2 to 5 where the overall
density of the lattice is unity and all the particles have unit mass.
}
\end{figure}

The density curves in the Figure correspond to an actual density of
unity.  The many crossings of the curves suggest that energetic flows
of highly inhomogeneous fluids would exhibit a complex structure
without any definite lattice structure while very slow flows might
``freeze'' into a least-energy crystalline form.  Simulations of the
Rayleigh-B\'enard problem (convection driven by a temperature gradient
in the presence of gravity) support this surmise\cite{b7}.
Low-energy, high-pressure simulations can actually ``freeze'', with
the smooth particles forming a locked lattice structure rather than
flowing.  For fluids this freezing behavior is undesirable. In the
next Section we consider the stress-free mechanical stability of the
three simplest two-dimensional lattice structures.

\section{Phase Instability from Density Potentials}
When discrete particles are involved there can be difficulties in
representing the smooth and continuous nature of fluid flows.  By
analogy with atomistic molecular dynamics, one
would expect that regular lattice arrangements of particles would
resist shear.  In the atomistic case in two space dimensions the
shear modulus $G$ is of the same order as the one-particle
Hooke's-law force constant evaluated from the Einstein model:
$$
G \simeq \kappa _{\rm Einstein} 
\equiv \frac{\partial ^2\Phi}{\partial x_1^2} \ ,
$$
where $x_1$ is the displacement of a single test particle, Particle
1, from its lattice site, with all the other particles fixed.  For a
sufficiently simple density-dependent  potential we can estimate the
one-particle force constant $\kappa  _{\rm Einstein}$ analytically.

Let us illustrate for the simplest possible density potential,
$$
\Phi _\rho \equiv \sum _j \frac{1}{2}[\rho _j - \rho_0]^2 \ ; \
\rho _j = \sum _i m_iw_{ij} \ ,
$$
where $\Phi _\rho $ is the total potential energy of the system,
$\rho _0$ is the target density minimizing that energy, and all
the particle masses are set equal to unity, $m_j = 1$.  The
first derivative,
$$
 \frac{\partial \Phi _\rho }{\partial x_1} =
\sum _j(\rho _1 + \rho _j - 2\rho _0)(xw'/r)_{1j} \ ,
$$
vanishes for
$$
x_1=0 \rightarrow \rho _1 = \rho _j = \rho _0 \ .
$$
The second derivative can be estimated by replacing the particle
sum with an integral:
$$
\frac{\partial ^2\Phi _\rho }{\partial x_1^2} = 
\sum _j(xw'/r)_{1j}^2 \simeq \int _0^h\left(\frac{xw'}{r}\right)^2
2\pi rdr = \frac{90}{7\pi h^4}\ .
$$
Fig. 2 shows that this analytic result closely resembles
the detailed lattice sums for all three regular two-dimensional
lattices.

\begin{figure}
\includegraphics{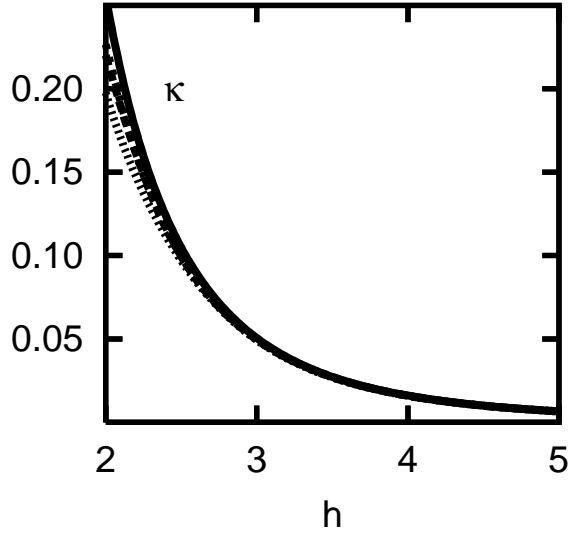}
\caption{
Comparison of the exact summed-up Einstein force constant
$\kappa _{\rm Einstein}$ with the
approximate integrated estimate as a function of the range $h$ of
Lucy's weight function.  The top-to-bottom ordering of the curves
is [integrated $>$ triangular $>$ square $>$ hexagonal].
}
\end{figure}

Nevertheless, our detailed investigation of this particular choice of
fluid model,
$$
E = \sum \frac{1}{2}(\rho - \rho _0)^2 \longleftrightarrow 
P = \rho ^2 (\rho - \rho _0) \ ,
$$
revealed that this expectation of a shear strength
varying as $h^{-4}$ is unfounded.  Instead, regular
lattices, with a stress-free, density-based potential corresponding
to an athermal fluid constitutive relation, show no shear resistance
whatever!

Numerical investigation shows that the square, triangular, and
hexagonal lattices, arranged {\em at} the target density $\rho _0$,
are {\em all} unstable to small displacements.  This can be shown by using
lattice dynamics, elastic theory, or molecular dynamics.  In every
case the regular lattices are unstable to a variety of shear modes.

The perfect-crystal elastic constants\cite{b8,b9} for this potential
can be calculated by two
chain-rule differentiations of the potential $\Phi $ with respect to
the elastic strains:
$$
C_{11}V =
\frac{\partial ^2 \Phi }{\partial \epsilon _{xx}^2} \ ; \
C_{12}V =
\frac{\partial ^2 \Phi }{\partial \epsilon _{xx}\partial \epsilon _{yy}} \ ; \
C_{44}V =
\frac{\partial ^2 \Phi }{\partial \epsilon _{xy}^2} \ ;
$$
$$
\epsilon _{xx} = \frac{\partial u_x}{\partial x} \ ; \
\epsilon _{yy} = \frac{\partial u_y}{\partial y} \ ; \
\epsilon _{xy} = \frac{\partial u_x}{\partial y} 
               + \frac{\partial u_y}{\partial x} \ .
$$
Here $u(r) = (u_x,u_y)$ represents an infinitesimal displacement from the
perfect-lattice configuration.  The resulting elastic constants take
the form of lattice sums:
$$
C_{11}V = \sum _i\left(\sum _j [x^2(w'/r)]_{ij}\right)^2 \ ;
$$
$$
C_{12}V = \sum _i\left(\sum _j [x^2(w'/r)]_{ij}\right)
                 \left(\sum _j [y^2(w'/r)]_{ij}\right) \ ;
$$
$$
C_{44}V = \sum _i\left(\sum _j [xy(w'/r)]_{ij}\right)^2 \ ;
$$
$$
r_{ij} = \sqrt{x_{ij}^2 + y_{ij}^2} \ ; \
x_{ij} = x_i - x_j \ ; \ y_{ij} = y_i - y_j \ .
$$
For the square, triangular, and hexagonal lattices it is evident,
by symmetry, that $C_{11}$ and
$C_{12}$ are equal and that $C_{44}$ vanishes.  The
nonvanishing elastic constant $C_{11}=C_{12}$ is exactly half
the bulk modulus $B$.  The range-dependence $B(h)$ is shown in
Fig. 3 for all three lattice structures.

\begin{figure}
\includegraphics{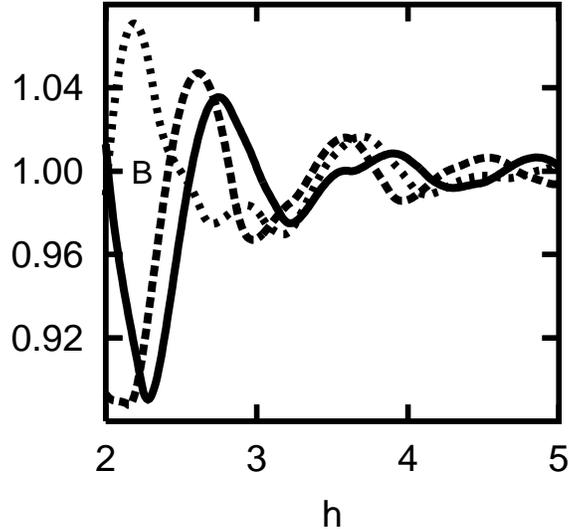}
\caption{
Variation of the bulk modulus $B = C_{11} + C_{12} = 2C_{11}
= 2C_{12}$ with the range of the weight function $h$ for three two-dimensional
lattices.  The top-to-bottom ordering of the curves at $h = 2.1$ is
[hexagonal $>$ square $>$ triangular]. 
}
\end{figure}

\section{Phase Stability from Density Gradients}
The results of the preceding section show that the smooth-particle
fluid model (correctly) is able to flow under an infinitesimal
shear stress.  For {\em solids} shear resistance is required.  A
simple potential supporting shear strength minimizes the gradient
of the density:
$$
\Phi _{\nabla \rho } \propto \sum _j \frac{1}{2}(\nabla \rho )_j^2 \ .
$$
This potential is minimized for regular lattices, in which there
can (by symmetry) be no density gradient {\em at} the particle
sites.  For systems with free surfaces---the details are not
considered here, but are elaborated in a forthcoming
book\cite{b10}---this potential also provides a surface
tension, eliminating the tendency of smooth particles to form
string-like phases.  Figs. 4 and 5 illustrate the stability
of the hexagonal lattice in the absence, and in the presence,
respectively of the density-gradient potential.  In the one example
detailed here (which is typical of many we have investigated,
with various sizes, initial conditions, and crystal structures)
the individual particle trajectories with and without the 
density-gradient potential are shown.  Evidently, by choosing
the proportionality constants wisely, these potentials can be
tuned to reproduce desired flow stresses for solids modelled
with SPAM.  This approach avoids many of the difficulties involved
in integrating the smooth-particle equations for the stress rates,
($\{ \dot \sigma \} \rightarrow \{ \sigma \} $).

\begin{figure}
\includegraphics{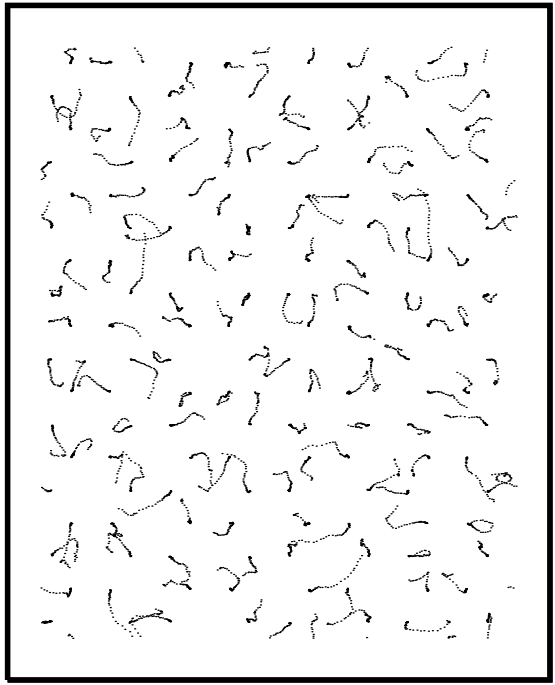}
\caption{
Hexagonal-lattice particle trajectories {\em without} the density-gradient
potential.  Initially the particle displacements were chosen randomly,
with zero sum and with an initial rms value of
$\sqrt{\langle \delta r^2 \rangle } = 0.02.$  The range of
Lucy's weight function is $h=3$ with the density and the particle
mass both chosen equal to unity.  The elapsed time (40,000 Fourth
Order Runge-Kutta timesteps $dt= 0.05$) is about 80 Einstein 
vibrational periods ($\kappa _{\rm Einstein} \simeq 0.05$).
$\Phi _\rho = \frac{1}{2}\sum (\rho - \rho _0)^2$.
}
\end{figure}

\begin{figure}
\includegraphics{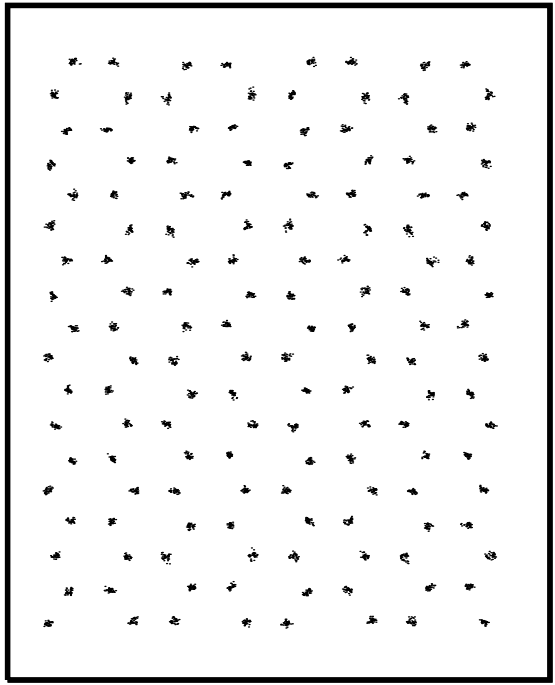}
\caption{
Hexagonal-lattice particle trajectories {\em with} the density-gradient
potential $\Phi _{\nabla \rho } = \frac{1}{2}\sum (\nabla \rho )^2$.
Initial conditions and length of the simulation are identical to those
of Figure 4.
$\Phi = \Phi _\rho + \Phi _{\nabla \rho } =
\frac{1}{2}\sum [(\rho - \rho _0)^2 + (\nabla \rho )^2]$.
}
\end{figure}

\section{Conclusions}
Density-dependent potentials can be used to simulate the behavior
of either fluids or solids from the standpoint of smooth-particle
simulation.  By introducing density-gradient potentials strength
and surface tension can be introduced, providing a useful model
for solids.  We believe that this idea will prove fruitful in a
wide variety of high-strain-rate applications of smooth-particle
methods.

\begin{acknowledgments}
Much of this work was carried out with the help of Chris Clark
and the support of the Academy of Applied Science's ``Research
in Engineering Apprenticeship Program'' at Great Basin College's
High Tech Center during the summer of 2005.  Some of the work
was performed under the auspices of the United States Department
of Energy at the Lawrence Livermore National Laboratory under
Contract W-7405-Eng-48.  We are specially grateful to Mike
MacFarlane (Great Basin College) and Bob Ferencz (LLNL) for
their help.
\end{acknowledgments}

\end{document}